\documentclass[fleqn,10pt]{wlscirep}
\usepackage[utf8]{inputenc}
\usepackage[T1]{fontenc}
\usepackage[acronym]{glossaries}
\usepackage{multirow}
\newcommand{\specialcell}[2][c]{%
  \begin{tabular}[#1]{@{}l@{}}#2\end{tabular}}
  
\title{Fused Audio Instance and Representation for Respiratory Disease
Detection}

\author[1,*]{Tuan Truong}
\author[1]{Matthias Lenga}
\author[2]{Antoine Serrurier}
\author[1]{Sadegh Mohammadi}
\affil[1]{Bayer AG, Berlin, Germany}
\affil[2]{Clinic for Phoniatrics, Pedaudiology and Communication Disorders, University Hospital of RWTH Aachen, Aachen, Germany}

\affil[*]{tuan.truong@bayer.com}

\newacronym{ml}{ML}{Machine Learning}

\newacronym{dl}{DL}{Deep Learning}

\newacronym{ai}{AI}{Artificial Intelligence}

\newacronym{mlp}{MLP}{Multilayer Perceptron}

\newacronym{cnn}{CNN}{Convolutional Neural Network}

\newacronym{vit}{ViT}{Vision Transformer}

\newacronym{covid19}{COVID-19}{Coronavirus Disease 2019}

\newacronym{relu}{ReLU}{Rectified Linear Unit}

\newacronym{dft}{DFT}{Discrete Fourier Transform}

\newacronym{stft}{STFT}{Short-Time Fourier Transform}

\newacronym{corona}{SARS‑CoV‑2}{Severe Acute Respiratory Syndrome CoronaVirus 2}
\newacronym{mfcc}{MFCC}{Mel Frequency Cepstral Coefficients}

\newacronym{zcr}{ZCR}{Zero-crossing Rate}

\newacronym{knn}{KNN}{K-Nearest Neighbors}

\newacronym{svm}{SVM}{Support Vector Machine}

\newacronym{copd}{COPD}{Chronic Obstructive Pulmonary Disease}

\newacronym{rnn}{RNN}{Recurrent Neural Network}

\newacronym{lstm}{LSTM}{Long Short-Term Memory}

\newacronym{miac}{MIAC}{\textbf{M}ulti-\textbf{i}n\textbf{s}tance \textbf{A}udio \textbf{C}lassification}

\newacronym{fair}{FAIR}{\textbf{F}used \textbf{A}udio \textbf{I}nstance and \textbf{R}epresentation}

\newacronym{auc}{AUC}{Area Under the Receiver Operating Characteristic  Curve}

\begin{abstract}
Audio-based classification techniques on body sounds have long been studied to aid in the diagnosis of respiratory diseases. While most research is centered on the use of cough as the main biomarker, other body sounds also have the potential to detect respiratory diseases. 
Recent studies on \acrshort{covid19} have shown that breath and speech sounds, in addition to cough, correlate with the disease. Our study proposes \textbf{\acrfull{fair}} as a method for respiratory disease detection. \acrshort{fair} relies on constructing a joint feature vector from various body sounds represented in waveform and spectrogram form. We conducted experiments on the use case of \acrshort{covid19} detection by combining waveform and spectrogram representation of body sounds. Our findings show that the use of self-attention to combine extracted features from cough, breath, and speech sounds leads to the best performance with an \acrfull{auc} score of 0.8658, a sensitivity of 0.8057, and a specificity of 0.7958. Compared to models trained solely on spectrograms or waveforms, the use of both representations results in an improved AUC score, demonstrating that combining spectrogram and waveform representation helps to enrich the extracted features and outperforms the models that use only one representation.
\end{abstract}
\begin{document}

\flushbottom
\maketitle
%
%
\thispagestyle{empty}

\section*{Introduction}
\label{introduction}
%
%
The human body produces numerous sounds that indicate its state of health. A slight change in the physical state may impact the way an organ operates and can consequently lead to irregular sound patterns. For instance, snoring is a common sound produced by upper airway obstruction during sleep. Snoring alone is generally not considered pathological, but if coupled with breathing pauses, it can be the manifestation of obstructive sleep apnea. More generally, body sounds can be used extensively to support diagnostic decisions. In particular, auscultation is a common technique used by clinicians to listen to internal sounds of the body with a stethoscope. Abnormal patterns in organs such as the heart, the lungs, and the gastrointestinal system can be detected using this method. In respiratory diseases such as pneumonia, auscultation can be performed to look for crackles or tubular breath sounds, an indication of pulmonary consolidation. Hence, body sound analysis is part of automated diagnostic applications such as in respiratory diseases \cite{song_diagnosis_2015, laguarta_covid-19_2020,botha_detection_2018,altan_deep_2020}, Parkinson's disease \cite{zhang_pdvocal_2019}, and sleep apnea \cite{kalkbrenner_apnea_2018}. Although detecting irregular internal sounds might be insufficient for a definitive diagnosis, it serves as an important indicator that can be combined with other clinical tests from different diagnostic tools to reach a conclusive diagnostic decision.

In this article, we study an audio-based approach for screening respiratory diseases. We focus on \acrfull{covid19}, a disease caused by \acrshort{corona} that infects the respiratory tract \cite{astuti_severe_2020}. Infected individuals express flu-like symptoms, making it difficult to differentiate between \acrshort{covid19} and other respiratory illnesses.  Viral testing through nucleic acid tests such as polymerase chain reaction (PCR) is a gold standard for diagnosis but takes several hours or even days to deliver results. Additionally, PCR testing requires specialized personnel and equipment that may not be available in low-income or remote areas. An alternative test, known as the antigen test, can retrieve results in less than 30 minutes by identifying viral proteins with specific antibodies. It is a viable option for mass testing but is less sensitive. Meanwhile, since \acrshort{corona} infects mainly the respiratory systems, it induces changes in some body sounds such as voice and breath, impacted by possible dysphonia and breath abnormalities, or creates sounds, such as cough. Several studies show that these changes are specific to \acrshort{covid19}. For example, a study by Huang et al.\cite{huang_respiratory_2020} finds abnormal breathing sounds in all \acrshort{covid19} patients. The irregular sounds include crackles, asymmetrical vocal resonance, and indistinguishable murmurs. In a different study of vocal changes in \acrshort{covid19} individuals \cite{al_ismail_detection_2021}, the authors validate the hypothesis that abnormal vocal fold oscillations are correlated with \acrshort{covid19}, inducing not only changes in voice but also the inability to speak normally. Body sounds, therefore, have the potential to serve as standalone or in parallel with antigen tests to detect \acrshort{covid19}.

Screening \acrshort{covid19} using body sounds has several advantages. Firstly, because PCR testing capacities are limited, screening with body sound or in conjunction with antigen tests can help prioritize who is eligible for PCR tests.  Allowing anyone with flu-like symptoms to order a PCR test will quickly exceed the testing capacity. Only suspects indicated by body sound screening could proceed with PCR tests. Body sound screening can rapidly identify suspect cases without asking them to quarantine while waiting for PCR results. Secondly, like antigen tests, body sound screening is fast, affordable, and convenient, and can be conducted without medical professionals. The cost of running body sound screening can even be lower than that of antigen tests because it can be installed as software or a mobile application on any device and uses the device microphone. Users do not need to buy additional support kits and can use their devices to record, analyze and monitor their status unlimited times. This is particularly useful in regions or countries where testing capacities are scarce, inaccessible, or expensive. Lastly, compared to antigen tests, it does not lead to medical waste because no physical products are manufactured, which alleviates the burden on the environment.

Given these advantages, the potential of body sounds for screening \acrshort{covid19} is enormous. However, a fully developed screening system using body sounds is not yet available. Current research on \acrshort{covid19} detection considering multiple body sounds often focuses on individual sounds and does not consider their interaction \cite{shimon_artificial_2021,suppakitjanusant_identifying_2021}. We hypothesize on the contrary that the effects of \acrshort{covid19} may occur in different body sounds or in a different combination of them, for different individuals. One or more body sounds may be affected, while the others remain intact. It is thus sensible not to rely on a single one but rather on a combination of several body sounds. We propose combining the most meaningful body sounds that are indicative of \acrshort{covid19} expressed in terms of fusion rules within the detection algorithm. Our hypothesis is stated as follows:
The cough, breath, and speech sounds contain biomarkers that are indicative of \acrshort{covid19} and can be combined using an appropriate fusion rule to maximize the chances of correct detection. To this end, we propose self-attention as a fusion rule to combine features extracted from cough, breath, and speech sounds. Mainly, we use waveforms and spectrograms as input to our model. A waveform represents an audio signal in the time domain, whereas a spectrogram is a representation in the time-frequency domain.
Our main contributions in this work are summarized as follows:
\begin{itemize}
    \item We demonstrate that cough, breath, and speech sounds can be leveraged to detect \acrshort{covid19} in a multi-instance audio classification approach based on self-attention fusion. Our experimental results indicate that combining multiple audio instances exceeds the performance of single instance baselines. 
    \item We experimentally show that an audio-based classification approach can benefit from combining waveform and spectrogram representations of input signals. In other words, inputting the time- and frequency-domain dual representations into the network allows for a richer latent feature space, which finally improves the overall classification performance.
    \item We integrate the above contributions into the \textbf{\acrshort{fair}} approach, a method that combines multiple instances of body sound in waveform and spectrogram representations to classify negative and positive \acrshort{covid19} individuals. The \textbf{\acrshort{fair}} approach is a general concept that can be applied to other sound classification tasks such as clinical problems related to other respiratory diseases.
\end{itemize}
\section*{Related work}
\label{related-work}
We briefly present the related work in body sound analysis for pulmonary diseases with a primary emphasis on the \acrshort{covid19} use case. Before \acrshort{covid19}, there is a well-established line of research on body sound analysis for pulmonary disorders such as tuberculosis, pneumonia, or \acrfull{copd}. Due to the urgency of the pandemic, this field of research has expanded and seen growing interest in newly developed techniques and collected datasets.
Most studies are centered on traditional machine learning techniques by building a classifier using extracted audio features of cough or respiratory sounds.
Botha et al.\cite{botha_detection_2018} study a combination of log spectral energies and \acrfull{mfcc} in screening tuberculosis using cough sounds of 38 subjects acquired in a specially designed facility. The authors achieve an accuracy of 0.98 and an \acrfull{auc} of 0.95 for the given task. 
Pahar et al.\cite{pahar_automatic_2021} investigate a similar task on cough sounds of 51 healthy and tuberculosis individuals in a primary healthcare clinic. The authors propose a linear regression model on extracted features, namely \acrshort{mfcc}, log spectral energies, \acrfull{zcr}, and kurtosis, which leads to a sensitivity and specificity of 0.93 and 0.95, respectively. 
Song\cite{song_diagnosis_2015} studies breath sounds to classify pneumonia among 376 children at three children's hospitals in Bangladesh. The author extracts a total of 18 acoustic features for the \acrfull{knn} and \acrfull{svm} classifier. The proposed method achieves 0.9198 accuracy, 0.9206 sensitivity, and 0.9068 specificity.
Antan et al.\cite{altan_deep_2020} investigate on multichannel lung sounds of 50 subjects from a multi-media respiratory database \cite{altan_multimedia_2017} to classify \acrshort{copd}. The authors develop a Deep Belief Network using features extracted using the Hilbert-Huang transform \cite{huang_empirical_1998}. The model achieves 0.9367 accuracy, 0.91 sensitivity, and 0.9633 specificity.
Xu et al.\cite{xu_listen2cough_2021} propose a multi-instance learning framework to process raw cough recordings and detect multiple pulmonary disorders including asthma and \acrshort{copd}. The presented framework achieves an F1-score of more than 0.8 in classifying healthy vs. unhealthy, obstructive vs. non-obstructive, and \acrshort{copd} vs. asthma. 
Kim et al.\cite{kim_accurate_2022} develop a deep learning approach to detect wheezing in children. The proposed framework consists of a ResNet34 model for spectrogram input and a \acrfull{mlp} for clinical data. The method has an \acrshort{auc} of 0.891 and F1 score of 0.872 on a private breathing audio dataset.
Petmezas et al.\cite{petmezas_automated_2022} propose a hybrid CNN-LSTM network to classify four types of lung sounds, namely normal, crackles, wheezes, and both crackles and wheezes. The model achieves state-of-the-art performance on the ICBHI 2017 dataset with an accuracy of 0.7369.
A recent and detailed review of disease classification from cough and respiratory sounds can be found in the work of Serrurier et al.\cite{serrurier_past_2022} and Xie et al.\cite{xia_exploring_2022}.

Regarding \acrshort{covid19}, there are large corpora of \acrshort{covid19} related audios collected from crowdsourcing. Voluntary participants submit recordings of their body sounds to a mobile app or website and provide metadata such as their \acrshort{covid19} status and comorbidity. Such large datasets enable researchers to develop \acrshort{covid19} detection algorithms as well as to benchmark their research work. To our knowledge, the largest crowdsourcing datasets are COUGHVID \cite{orlandic_coughvid_2021}, Coswara \cite{sharma_coswara_2020}, and Covid-19 Sounds \cite{brown_exploring_2020}. COUGHVID comprises more than 20000 cough recordings, while Coswara and Covid-19 Sounds consist of cough, breath, and vocal sounds from more than 2000 and 30000 participants, respectively. 
 In terms of technical development, a few studies follow the traditional machine learning approaches with handcrafted features\cite{fakhry_virufy_2021,meister_audio_2021,pahar_covid-19_2021,shimon_artificial_2021}. The most common audio features are still \acrshort{mfcc}, log Mel spectrogram, \acrshort{zcr}, and kurtosis. 
 Fakhry et al.\cite{fakhry_virufy_2021} propose an ensemble network of ResNet50 and \acrshort{mlp} on \acrshort{mfcc} and Mel spectrograms of cough recordings to classify \acrshort{covid19} individuals. The proposed solution claims an AUC of 0.99 on the COUGHVID dataset. 
 In a similar approach, the study by Meister et al.\cite{meister_audio_2021} benchmarks 15 audio features in the time and frequency domains for the \acrshort{covid19} detection task using cough and breath sounds. Their findings indicate that spectral features slightly outperform cepstral features in the classification task, and the best model is achieved using a \acrshort{svm} and Random Forest classifier, with AUCs of 0.8768 and 0.8778, respectively. 
 Several studies adopt Deep Learning approaches by training \acrfull{cnn} on spectrogram or waveform instead of handcrafted features.  
 Rao et al.\cite{rao_deep_2021} present a VGG13 network \cite{simonyan_very_2015} inputting spectrogram with combined cross-entropy and focal loss. The approach achieves an AUC of 0.78 on the COUGHVID dataset. 
 Xia et al.\cite{xia_covid-19_2021} provide an analysis of combined cough, breath and speech sounds using a simple VGG-ish model. The study introduces the combination of the features of various body sounds to improve classification performance. The best performance has an AUC of 0.75 and sensitivity and specificity of 0.70. 
 Wall et al.\cite{wall_deep_2022} put forward an ensemble approach by combining four deep neural networks with attention mechanism. the ensemble model is trained separately on respiratory, speech, and coughing audio inputs of the ICBHI and Coswara datasets. The overall performance for the base and ensemble model achieves ICBHI scores between 0.920 and 0.9766.
 Other studies also attempt pretraining on an external dataset or the same dataset without labels. The pretrained model is later finetuned on the target dataset with labels \cite{harvill_classification_2021, xue_exploring_2021, pinkas_sars-cov-2_2020}. 
 Harvill et al.\cite{harvill_classification_2021} pretrain all samples in COUGHVID dataset using autoregressive predictive coding with Long Short-Term Memory. The Mel spectrogram is split into several frames, and the model attempts to predict the next frame given the previous frames. The pretrained model is later finetuned on the DiCOVA dataset \cite{muguli_dicova_2021} and achieves an AUC of 0.95. 
 Similarly, Pinkas et al.\cite{pinkas_sars-cov-2_2020} pretrain a transformer-based architecture to predict the next frame of the spectrogram and transfers the pretrained features to a set of \acrshort{rnn} expert classifiers. The final prediction is the average of the scores produced by all expert classifiers. The proposed training scheme reaches a sensitivity of 0.78 on a private dataset collected by the authors. 
 Xue et al.\cite{xue_exploring_2021} propose use contrastive learning in a self-supervised pretraining phase. The contrastive pairs are created by randomly masking the inputs. The model is pretrained on the Coswara dataset without labels and finetuned with Covid-19 Sounds in the downstream task. The proposed technique achieves 0.9 AUC in the \acrshort{covid19} negative vs. positive classification task. 

Unlike research works that usually study each body sound independently \cite{suppakitjanusant_identifying_2021} or combined them by significant voting of prediction scores \cite{shimon_artificial_2021}, we explore fusion rules that combined them at the \textbf{feature level}. In other words, we train a network that learns a joint feature vector of all body sounds. Hence, the joint feature vector is optimized to implicitly reflect the relative importance of each body sound toward the final prediction. Although our work falls along the lines of Xie et al.\cite{xia_covid-19_2021}, we investigate a more complex fusion rule than simply concatenating features. We use self-attention \cite{vaswani_attention_2017}, which captures the dependencies among body sounds into a joint feature vector. Self-attention is used not only as a layer in the transformer architecture but can also be used to aggregate features \cite{truong_how_2021}. In addition, instead of using handcrafted audio features, we train our model using waveform and spectrogram representations, therefore creating more robust features compared to previous methods. We experiment our approach on the Coswara dataset and achieve state-of-the-art results. We report an average performance of the models obtained from cross-validation on a split test set in the \hyperref[cross-validation]{Method} section. However, we emphasize that there is no unique test set generated for the Coswara dataset and the data size was growing at the time we conducted our experiment. 
  
\section*{Methods}
\label{methods}
%
%
We present in this section our proposed approach: \acrfull{fair} for COVID-19 Detection. Let $\mathcal{D} = \left\{\mathbf{x}^{(i)}_j\right\}$ be a dataset of $n$ subjects, where $i \in \{1,...,n\}$ denotes the subject index and $j \in \{1,...,2c\}$ denotes the index of sound instances in the set of $c$ body sounds. The components $\mathbf{x}_1^{(i)},...,\mathbf{x}_c^{(i)}$ denote the fixed-length \emph{waveform} vectors related to the different $c$ audio instances. The components $\mathbf{x}_{c+1}^{(i)}, ...,\mathbf{x}_{2c}^{(i)}$ denote the associated \emph{spectrogram} representation of $c$ audio instances. The spectrogram is constructed by transforming the waveform representation with Discrete Short-Time Fourier Transform \cite{griffin_signal_1984}. In our experiments, we use the Mel-Spectrogram, which is the logarithmic transformation of the frequency in Hertz to Mel scale given by the equation:
\begin{equation}
    m = 1127\ln \bigg(1+\frac{f}{700}\bigg)
\end{equation}

Our objective is to derive a representative feature vector for $c$ body sounds per subject across waveform and spectrogram inputs.
We denote by $\mathbf{x}^{(i)} = [\mathbf{x}_1^{(i)},...,\mathbf{x}_{2c}^{(i)}]$ the aggregated input instance related to the $i$-th subject.
The \acrshort{fair} approach takes the input $\mathbf{x}^{(i)}$ of the $i$-th subject and returns a joint feature vector $\mathbf{z}^{(i)}$ that aggregates the information across multiple body sounds and representations as shown in the following equation.

\begin{equation}
\label{eq:fair}
    \mathbf{z}^{(i)} = \phi \left(\left[
    g_w \left(\left\{\mathbf{x}^{(i)}_1,..,\mathbf{x}^{(i)}_c\right\}\right), g_s \left(\left\{\mathbf{x}^{(i)}_{c+1},..,\mathbf{x}^{(i)}_{2c}\right\}\right)\right] \right)
\end{equation}

\begin{figure}[htp]
    \centering
    \includegraphics[width=\textwidth]{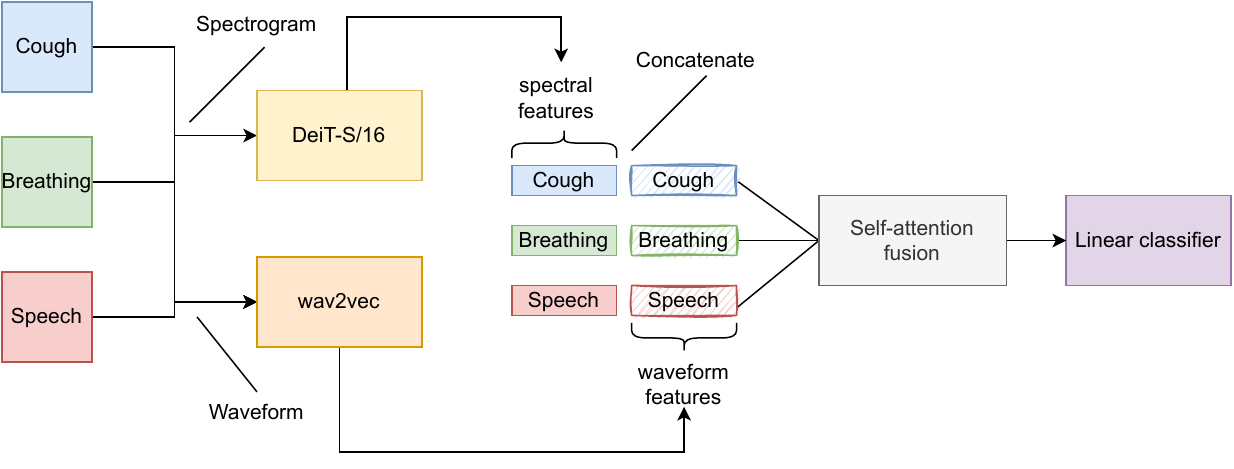}
    \captionof{figure}{An overview of \acrshort{fair} approach. Wav2vec and DeiT-S/16 extract waveform and spectrogram features from body sounds, which are then fused into a compact feature vector using self-attention. The resulting joint feature vector is used by the classifier, which is a \acrshort{mlp} that outputs the probability of \acrshort{covid19} infection.}
    \label{fig:hybrid}
\end{figure}

Here, $g_w$ and $g_s$ denote neural networks that extract features from waveform and spectrogram input, and $\phi$ is the attention-based fusion unit.
Figure \ref{fig:hybrid} shows an overview of the \acrshort{fair} approach and the main components along the pipeline.
The feature extractors and the fusion unit are instrumental components in our proposed approach and are further detailed in the next sections.
\subsection*{Feature extractors}
\label{feature-extractors}
Feature extractors are neural networks responsible for learning representative features for each body sound. As the input consists of waveform and spectrogram, two neural networks $g_w$ and $g_s$ are trained in parallel to handle both representations. In each network, the weights are shared across the input channels $1,.., c$ and $c+1, ..., 2c$. We choose $g_w$ to be a pretrained wav2vec \cite{baevski_wav2vec_2020} and $g_s$ to be DeiT-S/16, a \acrfull{vit} model \cite{dosovitskiy_image_2021}. DeiT-S/16 and wav2vec are transformer-based models and achieve state-of-the-art results in language and vision models. 

\begin{figure*}[htp]
    \centering
    \includegraphics[width=\textwidth]{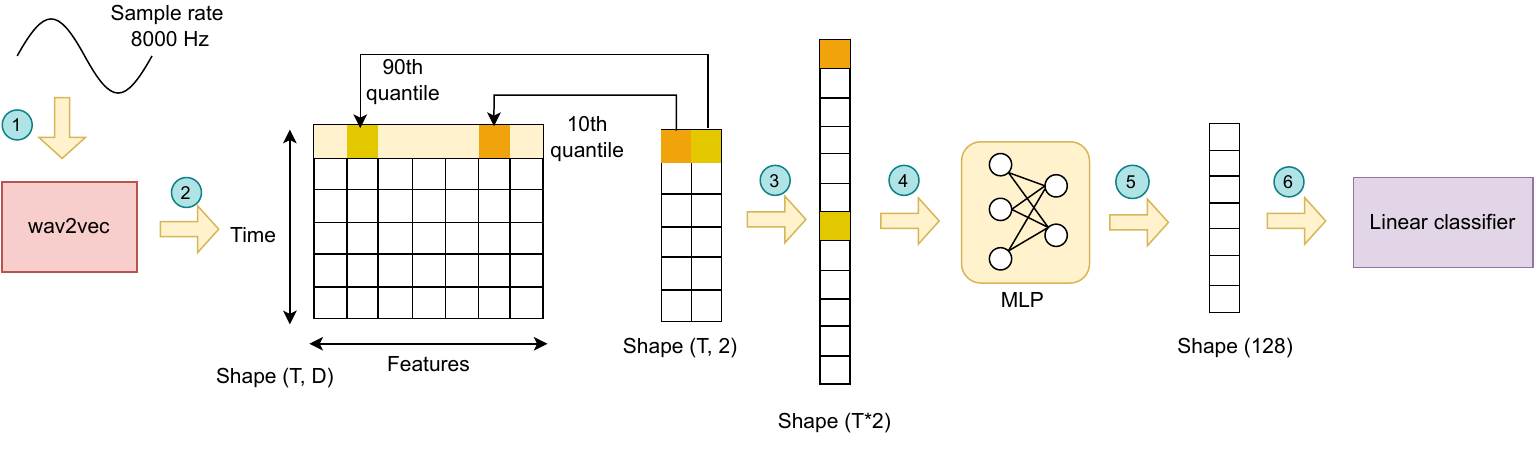}
    \caption[Wav2vec-based model for extracting waveform features]{Wav2vec-based model for extracting waveform features. (Step 1) 8 kHz sampled audio is fed into the pretrained wav2vec model to extract features. (Step 2) wav2vec outputs a feature vector per every 25 ms of the audio, resulting in a $t \times d$ matrix where $t$ is the total time indices and $d$ is the dimension of the feature vector. We select in each feature vector the element at the 0.1 and 0.9 quantile. (Step 3) The new feature matrix is flattened into a single vector. (Step 4) A \acrshort{mlp} layer receives the feature vector and (Step 5) projects it into a fixed dimension of 128.}
    \label{fig:waveform-extractor}

\end{figure*}

The wav2vec network \cite{baevski_wav2vec_2020} is developed to process audio for the speech-to-text translation task. It comprises both convolutional and self-attention layers and is pretrained on a large audio corpus in an unsupervised fashion. Therefore, we take advantage of the pretrained wav2vec features and designed a finetuning unit to effectively leverage them in our target dataset. As shown in Figure \ref{fig:waveform-extractor}, the recording is resampled to 8000 Hz, and features are extracted from every 25 ms using the frozen wav2vec unit. For each feature along the time axis, we select the values at the 0.1 and 0.9 quantiles, which can be considered to approximate the min and max pooling of feature vectors. The purpose of this step is to aggregate information over time by choosing only important information. After this step, we flatten the new feature matrix and feed it to a \acrshort{mlp} layer to reduce the dimension of the feature embedding to 128.

 The DeiT-S/16 architecture is a variant of \acrshort{vit} introduced by Touvron et al.\cite{touvron_training_2021} as part of the DeiT (Data-efficient image Transformers), which has the exact architecture of the original \acrshort{vit} \cite{dosovitskiy_image_2021} and differs only in the training strategy. The model is categorized into the small group where the projected embedding dimension through self-attention blocks is 384. It consists of 12 multi-headed self-attention layers, and each layer uses six heads. The resolution of each patch in the attention layer is 16 $\times$ 16 pixels. We change the last dense layer of DeiT-S/16 to be an identity unit so that we can extract the features from the previous layer. We use the pretrained DeiT-S/16 on ImageNet and finetune on our target dataset in all of our experiments. We projected the output unit with 128 feature vectors to have a fair comparison with wave2vec.

\subsection*{Fusion unit}
\label{fusion-unit}

We denote $\mathbf{f}_k^{(i)}$ with $k \in \left[1,c\right]$ the joint feature vector obtained by concatenating the feature vector of each individual body sound extracted from $g_w$ and $g_s$:

\begin{equation}
    \mathbf{f}_k^{(i)} = \left[g_w\left(\mathbf{x}^{(i)}_k\right), g_s\left(\mathbf{x}^{(i)}_{k+c}\right)\right]
\end{equation}

The fusion unit $\phi$ combines $\mathbf{f}^{(i)}_k$ with $k=1,..,c$ into a single vector $\mathbf{z}^{(i)}$ by using a multiheaded self-attention layer (MSA) \cite{vaswani_attention_2017} and a \acrshort{mlp} $h$:

\begin{equation}
    \mathbf{z}^{(i)} = \phi\left(\mathbf{f}^{(i)}_1,...,\mathbf{f}^{(i)}_c\right) = h\left(\text{MSA}\left(\mathbf{f}^{(i)}_1,...,\mathbf{f}^{(i)}_c\right)\right)
\end{equation}

The output of MSA for each subject $i$ is a new set of feature vectors $\left \{ \mathbf{f}'^{(i)}_k\right\}_{k=1}^c$ where each $\mathbf{f}'^{(i)}_k$ is a linear combination of original feature vectors $\left \{ \mathbf{f}^{(i)}_k\right\}_{k=1}^c$ weighted by the similarity score between feature $k$ and all $c$ features. Next, we concatenate all $\mathbf{f}'^{(i)}_k$ across $c$ vectors and feed the new concatenated vector into a \acrshort{mlp} layer $h$ to project it to the final 128-dimensional feature vector $\mathbf{z^{(i)}}$. The classifier takes $\mathbf{z}^{(i)}$ and outputs the predicted probability of whether the subject is infected with \acrshort{covid19} or not.

\subsection*{Dataset}
\label{experiments-dataset}
Coswara is a crowdsourcing project to build an audio corpus from \acrshort{covid19} negative and positive individuals. The dataset is publicly available to enable research on the development of diagnostic tools for respiratory diseases, in this case, \acrshort{covid19}. The dataset is published in the work of Bhattacharya et al.\cite{bhattacharya_coswara_2023} and publicly available at \url{https://github.com/iiscleap/Coswara-Data}. The approval of data collection was issued by the Institutional Human Ethics Committee at the Indian Institute of Science, Bangalore. The informed consent was obtained from all participants who uploaded the recordings. The collected data was anonymized and de-identified by the dataset's provider. The audio recordings were collected between April 2020 and September 2021 through crowdsourcing. 
Data collection occurs through a web interface where users are prompted to provide their metadata and recordings using a device microphone. The metadata covers age, sex, location and \acrshort{covid19} status. Users are then instructed to submit nine audio recordings of (heavy and shallow) cough, (deep and shallow) breath, (fast and slow) counting from 1 to 20, and uttering the phonemes /a/, /e/ and /o/. The \acrshort{covid19} status must be selected from the categories negative, positive with or without symptoms, recovered and not identified respiratory disease. There is no restriction on the duration of the recordings, so users can decide when they want to start and stop recording. We accessed the database when it was still in the last collection stage. All methods in our study were carried out in accordance with relevant guidelines and regulations. 

Regarding data preprocessing, we first remove the leading and trailing silence. We observe that long recordings ($>$20 seconds) mainly contain silence, and the duration at which people cough, breathe, or speak lasts only 3-10 seconds. 
Next, we remove corrupted files, which are those that contain no sound, noise, or a different sound type than the one reported in the label. The recordings whose duration is less than 1 second are eliminated because they do not contain any detected sound.
Then, similar to the approach of \cite{xia_covid-19_2021}, we use a pretrained model called YAMNet to systematically remove recordings where the detected sound is not the same as the provided label. YAMNet is a pretrained model on YouTube audio to classify 521 events, including cough, speech, and breath. If most of the predicted events in a recording are not cough, speech, or breath, we will remove all recordings associated with this participant. 
In addition, we decide not to use shallow cough and breath in our experiments because the quality of such recordings is low and can be misdetected as noise. After the process, 1359 participants are left, with 223 \acrshort{covid19} positive and 1136 \acrshort{covid19} negative. 
%
Each participant has exactly 7 recordings, which amounts to 9513 recordings used in our experiments. Table \ref{tab:dataset-statistics} provides statistics on the audio length. The participants are split into six folds for training and testing, and details are in \hyperref[cross-validation]{cross-validation} section.

\begin{table}[htp]
  \centering
  \small
  \begin{tabular}{l c c c c}
    \toprule
    \textbf{Body sound} & \textbf{Min (sec)} & \textbf{Max (sec)} & \textbf{Median (sec)} & \textbf{Mean (sec)}\\
    \midrule
    Heavy cough & 1.58 & 30.04 & 6.06 & 6.27 \\
    Deep breath & 2.65 & 30.04 & 16.30 & 17.08 \\
    Normal counting & 1.62 & 29.95 & 14.34 & 14.58 \\
    Fast counting & 1.86 & 29.95 & 7.94 & 8.00 \\
    Phoneme /a/ & 1.19 & 29.95 & 10.03 & 10.53 \\
    Phoneme /e/ & 1.28 & 29.95 & 10.96 & 11.73 \\
    Phoneme /o/ & 1.37 & 29.95 & 10.41 & 11.19 \\
    \bottomrule
  \end{tabular}
  \caption{The statistics of audio length (in second) after the preprocessing step.}
  \label{tab:dataset-statistics}
\end{table}

We use Torchaudio for audio processing and normalization. The values of loaded audio are automatically normalized between -1 and 1. We resample all recordings with two sample rates; 44100 Hz and 8000 Hz. The DeiT-S/16 uses a sample rate of 44100 Hz while the wav2vec model uses a sample rate of 8000 Hz. We decide to use only the four seconds of each recording, which is found by finetuning different lengths. In terms of spectrogram transformation, we take the Mel-spectrogram with 128 Mel filterbanks operating in 1025 frequency bins, \textit{i.e.}, FFT size of 2048, window size of 2048, and hop size of 1024.
We perform data augmentation on-the-fly during training. For each training audio, we randomly select a continuous 4-second interval out of the first 5 seconds of the recording to ensure a slight variation. However, during evaluation, we select the first 4 seconds in the audio. We investigate many audio augmentation techniques such as pitch shift, time stretch, or masking, but not all prove helpful in our tasks. Ultimately, only amplitude scaling, time and frequency masking are retained. In the amplitude scaling, we randomly inject an amplitude gain between 0.9 and 1.3 on the waveform. Amplitude scaling is always performed before spectrogram transformation in case the spectrogram representation is used. For the spectrogram, we additionally apply random time and frequency masking with a length of 10. 

\subsection*{Baseline and benchmark experiments}
\label{baseline-and-benchmark-experiments}
Our hypothesis states that the combination of body sounds can improve the detection of \acrshort{covid19} from audio recordings. Therefore, we compare the models developed with a single body sound instance, the baseline (BA) with multiple combinations of body sounds, the benchmark (BE). Table \ref{tab:overview-experiments} shows an overview of the baseline and benchmark experiments. In \textbf{baseline experiments}, we train seven models, each using only a single body sound instance and therefore without the fusion unit. The seven instances are heavy cough, deep breath, fast and normal counting, and the utterance of the phonemes /a/, /e/ and /o/. In \textbf{benchmark experiments}, we group counting and utterance of the three vowels as a single instance, thereafter speech. We investigate the following combinations: (1) speech, (2) cough and breath, (3) cough and speech, (4) breath and speech, and (5) cough, breath, and speech. In both baseline and benchmark experiments, we consider the input to be represented as either waveform or spectrogram in two separate experiments. The last experiment (BE3) is our \acrshort{fair} model, for which we use both waveform and spectrogram input. The input of DeiT-S/16 \cite{touvron_training_2021} is a spectrogram image of size 128 $\times$ 173 calculated from a 4-second audio clip sampled at 44100 Hz. The waveform input to wav2vec has a sample rate of 8000 Hz to be compatible with the pretrained wav2vec network. For 4 seconds, this waveform input vector corresponds to a length of 32000.

\begin{table*}[htp]
  \centering
  \small
  \begin{tabular}{l l l l l}
    \toprule
    \textbf{No.} & \textbf{Representation} & \textbf{Architecture} & \textbf{Body sound fusion} & \textbf{No. models}\\
    \midrule
    BA1 & Waveform    & wav2vec & None & 7\\
    BA2 & Spectrogram & DeiT-S/16 & None & 7\\
    BE1 & Waveform    & wav2vec & Attention & 5\\
    BE2 & Spectrogram & DeiT-S/16 & Attention & 5\\
    BE3 & \specialcell[t]{Spectrogram\\ Waveform} & \specialcell[t]{DeiT-S/16\\ wav2vec} & Attention & 5 \\
    \bottomrule
  \end{tabular}
  \caption{Baseline and benchmark experiments. The last experiment (BE3) is our proposed \acrshort{fair} model that uses both waveform and spectrogram inputs and the body sound fusion unit.}
  \label{tab:overview-experiments}
\end{table*}

\subsection*{Cross-Validation}
\label{cross-validation}
A set of 226 subjects (191 covid-19 negative and 35 positive), thereafter the test fold, is randomly selected from our data to serve as a fixed test set for all experiments. The remaining 1133 subjects are used as training and validation in a 5-fold cross-validation scheme as follows: the subjects are split into 5 folds of similar size (see Table \ref{tab:data-split}), 4 folds are used for training and the remaining fold for validation in a rotating process so that each subject is used exactly once as the validation fold. It provides five different models; each of them is tested on the fix test fold and the average of the results is reported in this article.

\begin{table*}[htp]
  \centering
  \small
  \begin{tabular}{c c c c c c c}
    \toprule
    \textbf{Subset} & \textbf{Label} & \textbf{Trial 1} & \textbf{Trial 2} & \textbf{Trial 3} & \textbf{Trial 4} & \textbf{Trial 5} \\
    \midrule
    \multirow{2}{5em}{Train} & Negative & 761 & 756 & 751 & 760 & 752 \\
    & Positive & 146 & 151 & 155 & 146 & 154 \\
    \midrule
    \multirow{2}{5em}{Validation} & Negative & 184 & 189 & 194 & 185 & 193 \\
    & Positive & 42 & 37 & 33 & 42 & 34 \\
    \bottomrule
  \end{tabular}
  \caption{Repartition of the subjects for the 5-fold cross-validation scheme}
  \label{tab:data-split}
\end{table*}

\subsection*{Hyperparameters} 
\label{hyperparameters}
\begin{table}[htp]
  \centering
  \small
  \begin{tabular}{l|c c|c c|c }
    \toprule
    \textbf{Architecture} & \multicolumn{2}{c|}{\textbf{wav2vec}} & \multicolumn{2}{c|}{\textbf{DeiT-S/16}} & \textbf{\acrshort{fair}}\\
    \hline
    \textbf{Body sound fusion} & None & Attention & None & Attention & Attention \\
    \textbf{Optimizer} & AdamW & AdamW & AdamW & AdamW & AdamW \\
    \textbf{Base learning rate} & 1e-4 & 1e-4 & 1e-4 & 1e-4 &1e-4 \\
    \textbf{Weight decay} & 1e-3 & 1e-3 & 1e-1 & 1e-1 & 1e-3\\
    \textbf{Optimizer momentum} & (0.9, 0.99) & (0.9, 0.99) & (0.9, 0.99) & (0.9, 0.99) & (0.9, 0.99) \\
    \textbf{Batch size} & 32 & 32 & 32 & 32 & 32  \\
    \textbf{Training epochs} & 30 & 30 & 30 & 30 & 30  \\
    \textbf{Learning rate scheduler} & cosine & cosine & cosine & cosine & cosine  \\
    \textbf{Warmup epochs} & 10 & 10 & 10 & 10 & 10  \\
    \textbf{Loss function} & BCE & BCE & BCE & BCE & BCE \\
    \bottomrule
  \end{tabular}
  \captionof{table}{Hyperparameter settings in baseline and benchmark experiments.}
  \label{tab:hyperparameters}
\end{table}

Table \ref{tab:hyperparameters} shows the complete hyperparameter settings in our experiments. Most hyperparameters are identical across architectures, representations, or fusion rules. For example, we train all models for 30 epochs without early stopping, and the best checkpoint is saved based on the best AUC obtained in the validation fold. The loss function that we use is binary cross-entropy (BCE), and we optimize this loss with AdamW (Adam with weight decay) \cite{loshchilov_decoupled_2019}. 
We fix a base learning rate of 0.0001 for all experiments and adjust the learning rate scheduler and weight decay conditional on the architecture or fusion rules. The weight decay factor is set between 0.1 and 0.001.

\subsection*{Evaluation}
\label{evaluation}
Our primary metric for model selection is \acrshort{auc}. During training, we save the checkpoint with the highest performance based on \acrshort{auc}. During validation, we use \acrshort{auc} to compute the optimal threshold and take this threshold to compute other metrics such as sensitivity and specificity in the test set. We report the \acrshort{auc} scores in the main paper and provide the sensitivity and specificity in the Appendix.

\section*{Results}
\label{results}

%
%
\subsection*{Baseline results}
\label{results-baseline}
Table \ref{tab:baseline-results} shows the performance of the models trained on a single body sound instance. The input to the model is either a waveform (BA1) or a spectrogram (BA2) of a single body sound. The results reveal that the models trained on spectrograms perform substantially better than those trained on waveforms. The average AUC scores for DeiT-S/16 (BA2) and wav2vec (BA1) are, respectively, 0.7549 and 0.6127. The performance of different body sounds across architectures and representations does not establish a consistent pattern. For example, using only cough sounds leads to the highest AUC score in DeiT-S/16, but a lower score in wav2vec. There appears to be a countertrend between DeiT-S/16 and wav2vec. For example, the counting sound achieves better results than the fast counting sound in DeiT-S/16 but worse in wav2vec. Similarly, the utterance of /o/ outperforms other vowels in DeiT-S/16 but performs poorly in wav2vec. 

\begin{table*}[htp]
  \centering
  \small
  \begin{tabular}{l c c}
    \toprule
    \textbf{Body sound} & \textbf{wav2vec (BA1)} & \textbf{DeiT-S/16 (BA2)} \\
    \midrule
    Cough - heavy    & .4574 $\pm$ .0093 & \textbf{.7782 $\pm$ .0132} \\
    Breath - deep & .6597 $\pm$ .0222 & \textbf{.7552 $\pm$ .0254} \\
    Counting - fast  & .7090 $\pm$ .0136 & \textbf{.7291 $\pm$ .0196}\\
    Counting - normal & .6285 $\pm$ .0155 & \textbf{.7943 $\pm$ .0326}\\
    Phoneme /a/ & .6484 $\pm$ .0150 & \textbf{.7418 $\pm$ .0399} \\
    Phoneme /e/ & .6209 $\pm$ .0197 &  \textbf{.7399 $\pm$ .0318} \\
    Phoneme /o/ & .5649 $\pm$ .0293 &  \textbf{.7457 $\pm$ .0288} \\
    \midrule
    Average          & .6127 $\pm$ .0751 & \textbf{.7549 $\pm$ .0215} \\
    \bottomrule
  \end{tabular}
  \caption{AUC scores of the baseline experiments for spectrogram (DeiT-S/16) and waveform (wav2vec) models. The bold scores denote the highest performance between spectrogram and waveform.}
  \label{tab:baseline-results}
\end{table*}

\subsection*{Benchmark results}
\label{results-benchmark}
Table \ref{tab:benchmark-results} shows the results for the \acrshort{fair} model (BE3) compared to the DeiT-S/16 (BE2) and wav2vec (BE1) models in different combinations of body sounds using self-attention fusion. In general, the \acrshort{fair} approach significantly outperforms models trained on a single representation. The average AUC score of \acrshort{fair} is 0.8316, which is 0.0227 more than DeiT-S/16 and 0.0847 more than wav2vec. \acrshort{fair} achieves the highest AUC scores in all combinations of body sound with the only exception in the cough-breath combination, which will be discussed in the next section. The cough-breath combination results in the lowest AUC score in all alternatives in terms of the body sound combination. The largest combination, cough-breath-speech, gives the best results in \acrshort{fair} and wav2vec but is behind the cough-speech combination in DeiT-S/16 by a margin of AUC 0.007. \acrshort{fair} achieves the highest AUC score of 0.8658 with the combination of cough, breath, and speech. This score is 0.0343 and 0.0941 higher than the best scores produced by DeiT-S/16 and wav2vec. The results of the \acrshort{fair} models find clear support for the use of dual audio representation along with body sound fusion.

\begin{table*}[htp]
      \centering
      \begin{tabular}{l c c c}
        \toprule
        \textbf{Model} & \textbf{wav2vec (BE1)} & \textbf{DeiT-S/16 (BE2)} & \textbf{\acrshort{fair} (BE3)} \\ 
        \midrule
        Speech          &.7562 $\pm$ .0152 & .8081 $\pm$ .0239 & \textbf{.8434 $\pm$ .0290} \\
        Cough + Breath  & .6739 $\pm$ .0435& \textbf{.7685 $\pm$ .0183} & .7585 $\pm$ .0174 \\
        Cough + Speech  & .7644 $\pm$ .0088& .8315 $\pm$ .0306 & \textbf{.8584 $\pm$ .0308} \\
        Breath + Speech & .7682 $\pm$ .0149& .8122 $\pm$ .0125& \textbf{.8319 $\pm$ .0187} \\
        Cough + Breath + Speech &.7717 $\pm$ .0128 & .8241 $\pm$ .0266 & \textbf{.8658 $\pm$ .0115} \\
        \midrule
        Average & .7469 $\pm$ .0369 & .8089 $\pm$ .0218 & \textbf{.8316 $\pm$ .0384} \\
    
        \bottomrule
      \end{tabular}
  \caption{AUC of the benchmark experiments for the spectrogram model (DeiT-S/16), the waveform model (wav2vec) and the \acrshort{fair} framework. The bold scores denote the highest performance between spectrogram, waveform and \acrshort{fair}.}
  \label{tab:benchmark-results}
\end{table*}

\section*{Discussion}
\label{discussion}

\begin{figure}[htp]
    \centering
    \includegraphics[width=0.6\linewidth]{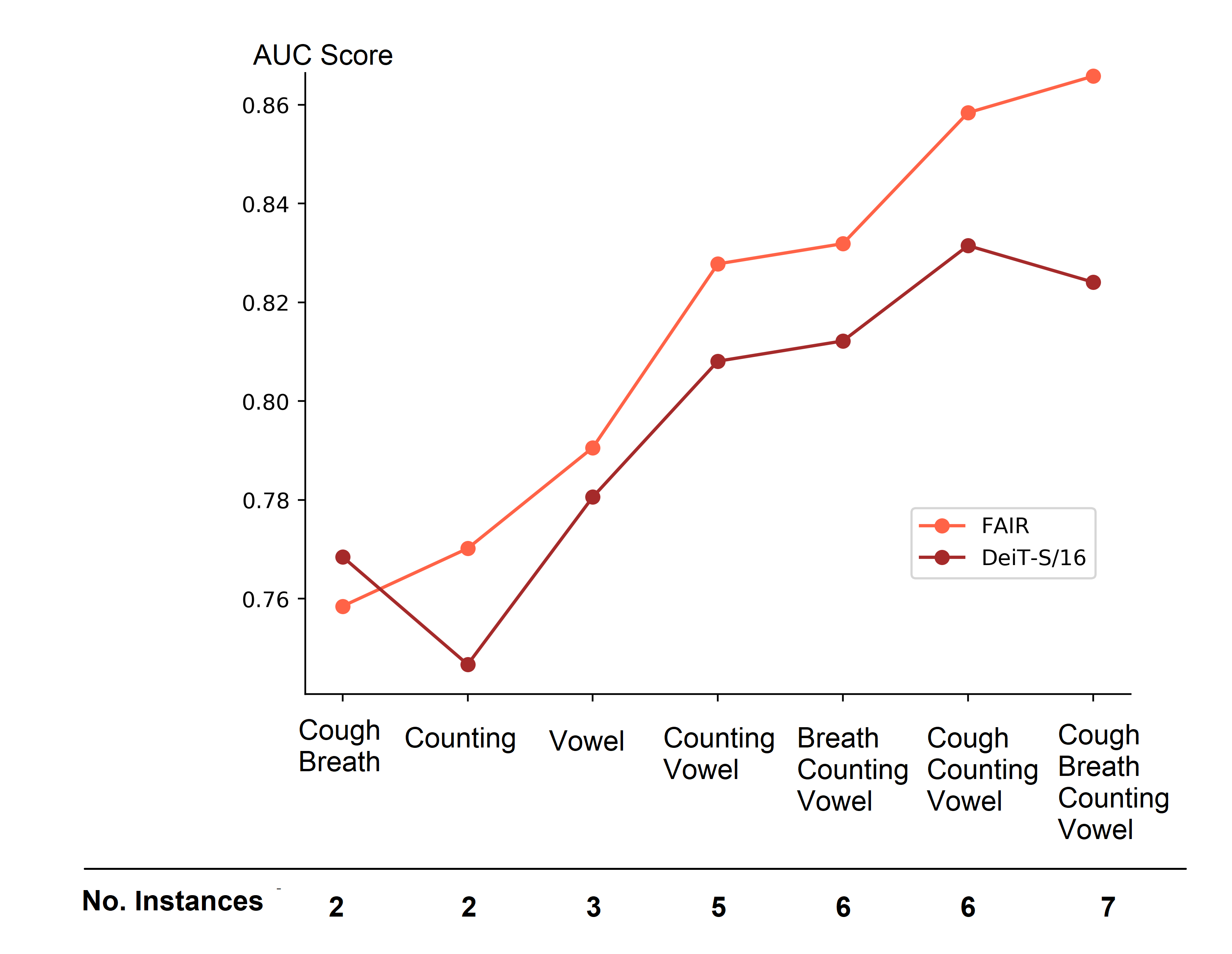}
    \caption{AUC scores of the DeiT-S/16 model and \acrshort{fair} framework vs. number of instances in each combination. The x-axis shows the combination with the number of instances in the ascending order of quantity. Additional results can be found in Table 1.2 and 1.4 of the supplementary material.}
    \label{fig:sortedAUC}
\end{figure}

As can be seen in Table \ref{tab:benchmark-results}, the AUC scores vary among body sound combinations, making it unclear which combination is best. Therefore, it is valid to doubt whether there is a preferable combination of body sounds that leads to the best predictive outcome. However, neither our results nor the literature provides a conclusive answer. We suggest that performance is correlated with the number of body sounds in a combination. To illustrate, we compare the performance of the model trained with (1) a single body sound instance and (2) a combination of body sounds.
In training models with a single body sound instance as input (\hyperref[results-baseline]{baseline results} section), no single body sound consistently outperforms the others.
The best-performing sound depends on the architecture or audio representation used. For instance, cough sound performs well with DeiT-S/16 (BA2) but not with wav2vec (BA1). Similarly, in our ablation study, replacing DeiT-S/16 with ResNet50 yields similar results (Table 1.3 in supplementary material). These subtle differences among body sounds may be due to the stochasticity or the feature extractor settings, indicating that no body sound is significantly better than the others as input to our model.
Regarding the combinations of body sounds (\hyperref[results-benchmark]{benchmark results} section), we observe that the combination of cough and breath consistently yields the lowest AUC scores for all models. This combination involves only two body sound instances, while all others include at least five instances. This observation suggests that the performance is likely to correlate with the number of body sound instances. To support this, we conduct additional experiments in a similar setting to benchmark experiments with the following combinations; counting (incl. fast and normal counting) and vowel (incl. utterance of /a/, /e/ and /o/). Figure \ref{fig:sortedAUC} shows that counting and cough-breath combinations perform similarly, while the three vowel utterances outperform the 2-instance combinations by 0.03-0.04 AUC. This supports a correlation between performance and the number of body sounds.

In addition to the number of body sounds in each combination, the varying duration of each instance can influence the results. In this study, we truncate each recording at 4 seconds. However, a body sound such as cough could last less than 4 seconds and the rest of the audio be just breath. A finer analysis taking this aspect into account should be considered in a follow-up study.

We analyze the effect of the dual representation of the spectrogram and waveform in the absence of body sound fusion by conducting an ablation study similar to the \acrshort{fair} framework but with the input of a single body sound. As there are no rules for body sound fusion, the features extracted from two representations are concatenated, flattened, and then projected onto a 128-dimensional vector by a \acrshort{mlp} layer. Similar to the baseline experiment, we present the AUC scores of 7 models trained on 7 body sound instances in Table \ref{tab:fair-baseline}. Overall, the average AUC scores are on par with those of the DeiT-S/16 model (BA2) in Table \ref{tab:baseline-results}. Breath and counting sound achieve the highest AUC score, whereas the utterance of vowel /e/ and /o/ leads to the lowest performance. The benefit of joint features from dual representation is not observed because the change in the individual AUC scores of each body sound does not follow any pattern.
Compared to the DeiT-S/16 results in Table \ref{tab:baseline-results}, except for cough, the difference in performance is subtle. The result suggests that the waveform representation contributes little to the final classifier. The performance is indeed strongly influenced by the powerful DeiT-S/16 in the spectrogram representation, which eclipses the features obtained from the waveform. Therefore, we conclude that using dual representation in the absence of body sound fusion does not improve any performance. However, when the dual representation is used for body sound fusion, the extra information from multiple body sounds is picked up by the fusion unit and enriches the joint extracted feature. The fusion unit is able to amplify the aggregated information due to the self-attention mechanism.
One of the interesting properties of self-attention is scaling, which is discussed in the work of Dosovitskiy et al.\cite{dosovitskiy_image_2021}. The authors note that the performance of the transformer-based model could be scaled up in response to an increase in resolution of patches or number of blocks. This contrasts with convolutional networks, in which accuracy can reach saturation at a certain level of complexity. This scaling property explains why adding more body sounds leads to a steady increase in AUC scores. Adding more body sounds means adding more tokens and establishing stronger dependencies among them. When only two or three instances of body sound are adopted, the effect of body sound fusion is less significant. Figure \ref{fig:sortedAUC} shows the AUC scores of the \acrshort{fair} and DeiT-S/16 models on the different combinations of body sounds sorted in ascending order of instances. Combinations with less than or equal to three instances, \textit{i.e.}, cough-breath, fast and normal counting, /a-e-o/ vowel utterance achieve AUC scores in the range of 0.75-0.79, which is on par or slightly better than the performance of models on a single instance (Table \ref{tab:fair-baseline}). This happens because the number of instances is insufficient to establish long-range dependencies. As more body sounds are added, these dependencies are captured, and the performance of models with fusion units starts to improve substantially. 
We observe a similar effect when replacing the fusion unit of \acrshort{fair} by attention weighted pooling (see supplementary material). When the number of body sounds in the combination is less than three, both attention-based fusion units have comparable performance. However, the gap is significant as more instances are combined.
In addition, the joint feature vector embeds more information when a dual representation is adopted. When the number of instances in the combination is small, \textit{i.e.}, less than three, the gain due to the dual representation is not noticeable. However, starting from five instances, the gap between \acrshort{fair} and DeiT-S/16 becomes wider in favor of \acrshort{fair}. We attribute this gain to the resonance of extra information given by the dual representation and the number of body sounds, which efficiently captured the self-attention fusion rule.

  \begin{table}[htp]
    \centering
      \begin{tabular}{l c}
        \toprule
        \multicolumn{2}{c}{\textbf{Dual representation without fusion rules}} \\
        \midrule
        Architecture & DeiT-S/16 \& wav2vec \\ 
        \midrule
        Cough - heavy    & .7426 $\pm$ .0268  \\
        Breath - deep & .7661 $\pm$ .0113  \\
        Counting - fast  & .7698 $\pm$ .0204  \\
        Counting - normal& .7581 $\pm$ .0938  \\
        Phoneme /a/       & .7577 $\pm$ .0213  \\
        Phoneme /e/         & .7299 $\pm$ .0174  \\
        Phoneme /o/         & .7394 $\pm$ .0168  \\
        \midrule
        Average          & .7519 $\pm$ .0137 \\
    \bottomrule
  \end{tabular}
  \caption{Baseline performance (in AUC) of \acrshort{fair} on a single body sound.}
  \label{tab:fair-baseline}
  \end{table}

\section*{Conclusion}
\label{conclusion}

In this article, we study Deep Learning approaches to detect \acrshort{covid19} using body sounds. To this end, we propose \acrshort{fair}, a multi-instance audio classification approach with attention-based fusion on waveform and spectrogram representation. We prove the effectiveness of our approach by conducting extensive experiments on the Coswara dataset. The results demonstrate that the fusion of body sounds using self-attention helps extract richer features that are useful for the classification of \acrshort{covid19} negative and positive patients. In addition, we perform an in-depth analysis on the influence of the fusion rule on the performance. We found that the scaling property of self-attention shows great efficiency when more instances of body sounds and representations are adopted. The best setting with a combination of cough, breath, and speech sounds in waveform and spectrogram representation results in an AUC score of 0.8658, a sensitivity of 0.8057, and a specificity of 0.7958 on our test set. The sensitivity of our model exceeds 0.75, the required threshold of the \acrshort{covid19} screening test \cite{scheiblauer_comparative_2021}. 

In addition, \acrshort{fair} is not limited to \acrshort{covid19} detection. It can be adapted to other audio classification problems involving diverse combinations of multi-instance inputs.  In our future work, we consider applying \acrshort{fair} to other critical biomedical audio classification tasks. The framework can be extended in various ways, for example by integrating multi-modal inputs, such as clinical lab values, with the spectrogram and waveform features derived from the audio signal. The attention-based fusion mechanism allows quantifying the feature attribution based on the attention weights. Particularly in the multi-modal setting, we propose to carefully assess the aforementioned attribution scores in order to derive further insights into the relevance of different audio or non-audio clinical biomarkers. Furthermore, as indicated in Truong et al.\cite{truong_how_2021}, simultaneously extracting and fusing multiple multi-modal embeddings could the overall model performance in classification tasks by leveraging complementary information within an extended feature space.
\bibliography{sample}

\section*{Data availability}
The datasets analyzed during the current study are published in the work of Bhattacharya et al.\cite{bhattacharya_coswara_2023} and publicly available at \url{https://github.com/iiscleap/Coswara-Data}. The approval of data collection was issued by the Institutional Human Ethics Committee at the Indian Institute of Science, Bangalore. The informed consent was obtained from all participants who uploaded the recordings. The collected data was anonymized and de-identified by the dataset's provider. All methods were carried out in accordance with relevant guidelines and regulations.
\section*{Legend}
\paragraph{Figure \ref{fig:hybrid}} An overview of \acrshort{fair} approach. Wav2vec and DeiT-S/16 extract waveform and spectrogram features from body sounds, which are then fused into a compact feature vector using self-attention. The resulting joint feature vector is used by the classifier, which is a \acrshort{mlp} that outputs the probability of \acrshort{covid19} infection.
\paragraph{Figure \ref{fig:waveform-extractor}}Wav2vec-based model for extracting waveform features. (Step 1) 8 kHz sampled audio is fed into the pretrained wav2vec model to extract features. (Step 2) wav2vec outputs a feature vector per every 25 ms of the audio, resulting in a $t \times d$ matrix where $t$ is the total time indices and $d$ is the dimension of the feature vector. We select in each feature vector the element at the 0.1 and 0.9 quantile. (Step 3) The new feature matrix is flattened into a single vector. (Step 4) A \acrshort{mlp} layer receives the feature vector and (Step 5) projects it into a fixed dimension of 128.
\paragraph{Figure \ref{fig:sortedAUC}} AUC scores of the DeiT-S/16 model and \acrshort{fair} framework vs. number of instances in each combination. The x-axis shows the combination with the number of instances in the ascending order of quantity. Additional results can be found in Table 1.2 and 1.4 of the supplementary material.

\section*{Author contributions statement}
M.L and S.M conceived the presented idea and planned the experiment. T.T carried out the experiment and wrote the manuscript with support from M.L, A.S, and S.M. All authors provided critical feedback and helped shape the research, analysis and manuscript.

\section*{Additional information}
The author(s) declare no competing interests.
\end{document}